\documentclass[%
 reprint,
superscriptaddress,
 amsmath,amssymb,
 aps, floatfix
]{revtex4-2}
\usepackage{float}
\usepackage{graphicx}
\usepackage{amsmath,amsfonts, amssymb}
\usepackage{mathtools}
\usepackage[dvipsnames]{xcolor}
\usepackage{braket}
\usepackage{booktabs}
\definecolor{ProGreen}{RGB}{0, 200, 0}

\usepackage[colorlinks=true, citecolor=ProGreen, linkcolor=magenta, urlcolor=magenta]{hyperref}
\usepackage[capitalize]{cleveref}
\begin{document}

\title{Improving device-independent quantum key distribution protocols through multiple routed Bell tests}

\author{Sujan Vijayaraj}
\email{sujan.vijayaraj@unipa.it}
\affiliation{Universit\`a degli Studi di Palermo, Dipartimento di Fisica e Chimica - Emilio Segr\`e, via Archirafi 36, 90123 Palermo, Italy}

\author{Mauro Paternostro}
\affiliation{Universit\`a degli Studi di Palermo, Dipartimento di Fisica e Chimica - Emilio Segr\`e, via Archirafi 36, 90123 Palermo, Italy}
\affiliation{Centre for Quantum Materials and Technologies, School of Mathematics and Physics, Queen's University Belfast, BT7 1NN, United Kingdom}

\date{\today}

\begin{abstract}
Device-independent quantum key distribution (DI-QKD) guarantees security with a minimal set of assumptions about the experimental setup. The challenge posed by its implementation could be tackled using routed Bell tests with entanglement swapping, or distant Bell state measurement (BSM) units. However, practical distances still require local tests with close-to-ideal violations or BSMs with high efficiency. We propose a DI-QKD protocol based on a multi-port BSM, and multiple sources and measurement devices, where in each round, routed tests are performed on randomly selected local devices. The violation of local Bell tests is checked even for the subset of rounds where successful BSM projections are achieved. By requiring that this violation remain consistent with the overall average, we specifically show improvements in the critical efficiencies of a three-port BSM, of up to $6\%$ in absolute terms for high visibilities. Without ideal violations, we also achieve arbitrarily small efficiencies with arbitrarily many BSM ports. Our approach enables long-distance DI-QKD, with access to highly efficient loophole-free routing setups, and multiple local tests (possibly imperfect) with very high local detection efficiencies. Remarkably, it avoids the use of quantum memories and the need for long-range Bell tests. Finally, we extend the concept of routing to dimension witnesses, where qubit-bounded sources send states to the BSM. This can be seen as a semi-device-independent extension of the aforementioned protocol.
\end{abstract}

\maketitle

\section{Introduction}
Quantum key distribution (QKD) protocols offer information-theoretic security, by making certain assumptions about the underlying hardware. For more comprehensive security applications, device-independent (DI) QKD protocols consider uncharacterised devices and sources, with the smallest set of assumptions about the devices~\cite{zapatero2023advances}. The few -- sometimes implicit -- assumptions address the trusted and random nature of inputs, the memoryless nature of the measurement devices, and the trustworthiness of the local computers handling the protocol. The realization of DI-QKD protocols poses several problems, the need for long-range Bell tests between distant parties being one of the biggest challenges.

{\it Routed Bell tests} have become a viable technique for self-testing devices close to the source~\cite{chaturvedi2024extending}, while retaining the DI advantage of requiring minimal assumptions. They facilitate the easing of requirements on detection efficiency for long-distance devices, conditioned on the source demonstrating close-to-ideal local Bell test violations~\cite{leroydeloison2025device, tan2024entropy}. The only extra hardware requirement is about the switch performing routing: information on the switch's routing choice must be shielded from the devices and the source. Combined with Bell state measurements (BSMs), like in measurement-DI (MDI) QKD with two sources~\cite{lo2012measurement}, they can operate at arbitrarily long-range BSM efficiencies for ideal local tests \cite{kossmann2025routed}. Indeed, the first protocol using local Bell tests \cite{lim2013device} was inspired by MDI-QKD. However, this version of routed Bell tests (and prior proposals as explained in Ref.~\cite{lobo2024certifying}) can come with source loopholes that must be properly accounted for. 

When the local tests are imperfect, an adversary (referred to as Eve) can apply a convex strategy for the two sources (each for Alice and Bob), such that the actual key is generated by more local states than those observed by the test. This is made possible due to non-ideal efficiencies of the distant BSMs. The source strategy can be correlated to BSM events, such that ideal violations may be blocked, and more local rounds can be used selectively for successful BSM projections (part of key rounds). This can lead to an overestimation of secret key rates and  become a critical vulnerability in the DI scenario, given that the sources and the BSMs may be controlled by Eve. Outside of this BSM setup, previous works have shown the possibility of generating key with low detection efficiencies by increasing the number of measurement inputs~\cite{sekatski2025certification}, or using multi-outcome hyper-entangled states \cite{chaturvedi2025extending}. However, they still require long-range Bell tests and pose significant added difficulties in experiments. In event-ready setups, measurements are made only after successful BSM projections. This requires quantum memories, which have not been feasible for practical distances~\cite{zapatero2023advances}.

We focus on using two-qubit entangled sources with standard measurement settings and the standard Clauser–Horne–Shimony–Holt (CHSH) inequality \cite{clauser1969proposed}, by making use of BSM, as in Ref.~\cite{kossmann2025routed}. We solely deal with local tests, and avoid the use of quantum memories like in an event-ready setup for long-range Bell tests \cite{zapatero2023advances}. Our goal is to enable long-distance DI-QKD, only limited by the observed (possibly imperfect) local Bell tests, BSM efficiency and the quantum bit error rate (QBER). We show that the critical BSM efficiency can be drastically reduced by increasing the number of measurement devices, entanglement sources, and by deploying a multiport-BSM \cite{zukowski1997realizable}. With these additional resources, we achieve low critical efficiencies by simultaneously performing local Bell tests in every round, including when there is a successful BSM projection. We require that the observed Bell test conditioned on the rounds with successful BSM projection, match the overall observed Bell test. We will refer to this technique as {\it decoy Bell test} (DBT). By making the standard assumption that the routing path is unknown to the devices, DBTs can potentially limit the adversarial strategy that pushes more local states to key generation rounds. We use a simple linear program to estimate the new substantial CHSH violation for the resulting key rounds, with the possibility that the underlying source strategy can still be correlated to BSM announcements (success or failure in projection). We use this violation to then compute the asymptotic key rate. For example, when achieving a Bell-CHSH parameter of $2.77$, we show that a $3-$port BSM and $2$ DBTs can allow for BSM efficiencies down to $\sim 9.4\%$. Not using DBTs would allow efficiencies starting from approximately $15.2\%$. This is a significant reduction in critical efficiency, which can drastically improve the maximum achievable distance for current DI-QKD protocols. For $S \gtrsim 2.75$, we observe a reduction of up to $6\%$ (in absolute terms) in critical BSM efficiency requirements, using a $3-$port BSM.

Similar to the MDI-QKD protocol in architecture, we extend the routing strategy to qubit sources, making use of dimension-witnesses \cite{gallego2010device} close to the source, to ensure correct encoding of the dimension-bounded qubit source. The security is based on the same principle as the semi-device-independent (SDI) QKD protocol~\cite{pawlowski2011semi}.

The proposed protocols enable us to shift our focus to short-range Bell tests for long-distance experiments. The remainder of this paper is organized as follows. In Sec.~\ref{RoutedTests} we describe the routed Bell test protocol with BSMs. In Sec.~\ref{DBTProtocol}, we propose the modified protocol, based on DBTs. In Sec.~\ref{RoutedWitnesses}, we describe routed witnesses and the SDI protocol. 

\section{Routed Bell Tests}
\label{RoutedTests}
\subsection{Setup}
We consider memoryless, non-signalling and ideally shielded measurement devices. In particular for the switches, we assume that the sources and the key generating measurement devices (Alice and Bob), cannot learn the routing path. This is important for protocols based on routing \cite{lobo2024certifying}. We also assume unit detection efficiencies for the local devices.

\begin{figure}
    \centering
\includegraphics[width=0.75\linewidth]{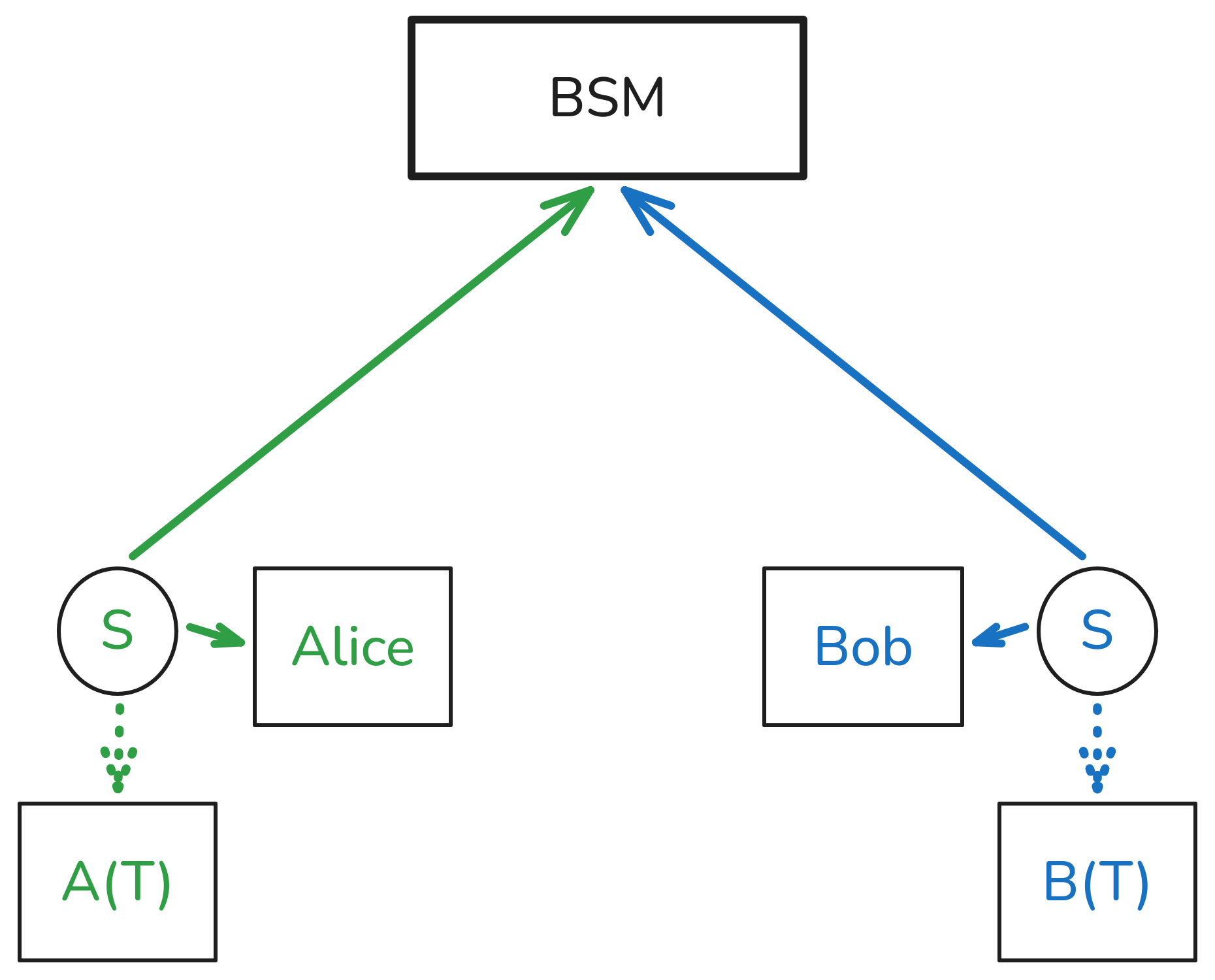}
    \caption{The routed Bell test with a distant BSM and local CHSH tests, as proposed in Koßmann et al. \cite{kossmann2025routed}, can be used to generate key using local tests. $A(T)$ and $B(T)$ are test devices, used exclusively for the local CHSH test on their respective sides. Alice and Bob receive a state in every round from their nearest sources. Unbeknownst to the sources and the measurement devices, the switch controls the path of the other qubit to the BSM or the local test.}
    \label{fig:routed_original}
\end{figure}

In what follows, we will consider the BSM version of routed Bell tests \cite{kossmann2025routed}, with several caveats, as shown in Figure \ref{fig:routed_original} and explained below. The BSM unit is responsible for publicly announcing the BSM outcomes $z\in \{0, 1\}$, where $z=0$ corresponds to an inconclusive result or a non-detection, and $z=1$ to the successful identification or projection. For the local tests, we use the standard two-party CHSH test \cite{clauser1969proposed}
\begin{equation}
S = \sum_{xx'}(-1)^{xx'}[E(a=a'|x,x')- E(a\neq a'|x,x')] \leq 2,
\end{equation}
where $x:\{0, 1\}$ denotes the choice of measurement inputs for Alice, and $x':\{0, 1\}$ for the local test device on Alice's side. The measurement outputs are $a:\{0, 1\}$ and $a':\{0, 1\}$ respectively. Additionally, we have another CHSH test on Bob's side, with Bob's inputs: $y$ and the local test device: $y'$. The outputs are similarly $b$ and $b'$ respectively. Let Alice's random bit, $r_A: \{0, 1\}$ (similarly, $r_B$ for Bob), control a switch that is used to route one qubit from her source to either destination. The state is mostly sent to the Bell state measurement (BSM) unit ($r_A=0$), but sometimes sent and measured close to the source ($r_A=1$). The key rounds are composed of $z=1$ rounds, when Alice and Bob use the same bases. The QBER $q$ can be calculated from the proportion of the key rounds where $a \neq b$ (after flipping in case of projection to anti-correlated states) as,
\begin{equation}\label{q}
q=   P(a \neq b | x=y, z=1). 
\end{equation}
The main parameter of non-ideal efficiency $\eta$ is the efficiency of the BSM ($0 < \eta\leq 1$), also equivalent to the proportion of rounds where $z = 1$. This gets much lower for BSM using heralded qubit amplifiers \cite{gisin2010proposal} in the context of routed Bell tests (typically $1\%$). Hence, they are not suitable for protocols with solely local Bell tests with non-ideal violation.
\subsection{Protocol}
We briefly state the protocol to establish the key between Alice and Bob.

1. Each source is expected to ideally produce a Bell state in each round.

2. Both Alice and Bob expect to receive a qubit from their nearest source, and use their random choice of inputs ($x, y$) to measure in each round. Their ideal inputs are $Z$ and $X$.

3. On each side, a switch routes the second qubit from the nearest source to either the BSM ($r_A=0$ for Alice, or $r_B=0$ for Bob), or the test device ($r_A=1$ for $A(T)$, or $r_B=1$ for $B(T)$) with ideal inputs being $(Z\pm X)/\sqrt{2}$.

4. After repeating the steps for several rounds as predetermined, the BSM projection (failure or success, denoted by $z=0$ or $z=1$ respectively) results are announced. 

5. Alice and Bob estimate the CHSH violation, from their respective local tests. If the violation is sufficient, they proceed. Otherwise, they abort the protocol.

6. Alice and Bob announce $x$ and $y$ for all rounds. Using classical communication, Alice and Bob estimate the QBER, $q$, from a subset of rounds where $z=1$ (and $r_A=r_B=0$) and $x=y$. If it is within an acceptable threshold, they proceed with error correction and privacy amplification, for rounds where $z=1$ (and $r_A=r_B=0$) and $x=y$. Otherwise, they abort the protocol.

\subsection{Security}
We use the fact that measurement in the local devices (with ideal detection efficiency) takes place immediately and a basis dependence is forced on the qubits sent to the distant BSM. This is equivalent to sending the BB84 states to the BSM like in the MDI-QKD protocol \cite{lo2012measurement}. Alice and Bob wait until $z$ is publicly announced after all rounds, and proceed with sifting only after, ensuring that $z$ is independent of $x$ and $y$.
In this setup, we start by noting that due to the linearity of the CHSH inequality, the observed violation, $S$, can be written as an adversarial convex combination in the following way,
\begin{equation}
  S = S^{z=0}(1-p_b\eta)+S^{z=1} p_b\eta,
\end{equation}
where $p_b$ is the probability with which both switches route the state to the BSM, and Eve sets $z=0$ and $z=1$ at the BSM, for violations corresponding to $S^{z=0}$ and $S^{z=1}$ respectively. This is the most general attack. Since rounds corresponding to $S^{z=1}$ (they occur simultaneously for Alice and Bob) are used for key generation, the optimal attack requires $S^{z=0}=2\sqrt{2}$. We can write
\begin{equation} \label{eq 1}
  S = 2\sqrt{2}(1-p_b\eta)+S_{k} p_b\eta,
\end{equation}
where we denote $S_k=S^{z=1}$, and $0<p_b<1$. We will refer to each component of the convex combination as a {\it strategy}. Since Eve can selectively block the BSM contributions from the ideal strategy (corresponding to  $S=2\sqrt{2}$), she can set the minimum $S_k$ for given $S$, $p_b$ and $\eta$. In general, in a given round, when the $S_{A_k}$ strategy [as in \cref{eq 1}] is exclusively chosen from Alice's device for $z=1$, where
\begin{equation}
\label{SaSb}
\begin{aligned}
    S_A=2\sqrt{2}(1-\eta)+S_{A_k}\eta,\\
    S_B=2\sqrt{2}(1-\eta)+S_{B_k}\eta,
\end{aligned}
\end{equation}
Bob's device contains the $S_{B_k}$ strategy in the same round, where we absorb the routing probability by setting $p_b = 1$ ($p_b$ can be included as a multiplicative factor to $\eta$ if required). We assume from now on that $S_{A,B}=S$. Hence, the key rounds are exclusively made up of our new CHSH constraints: $S_{A_k}=S_{B_k}=S_k$. 

To assess the security of the protocol, we use the one-way (from Alice to Bob) Devetak-Winter secret key rate, $R \geq H(A|E)-H(A|B)$~\cite{devetak2005distillation}, where $H(A|B)=h(q)$. Establishing a lower bound for $H(A|E)$ from the constraints is the main problem, and we follow the exact procedure highlighted in Ref.~\cite{kossmann2025routed}. For Alice and Bob, we need to compute the $\inf \{H(A|E)\}$ for the purification $\ket{\psi}_{ABE}$ such that the marginal constraints for each party's system, observed $S$, global QBER, and general optimization constraints hold. 

In the current protocol, no key rate can be obtained for $\eta \lesssim 15.2\%$ when $S=2.77$, or $\eta \lesssim 1.85\%$ when $S\approx2.817$. It remains of practical interest to obtain positive key rate at lower BSM efficiencies for non-ideal violations. Naturally, it would be useful to devise a protocol where $S_{A_k} \approx S$ for any $0< \eta < 1$. We will show that this is possible with multiple local Bell tests. 

\section{Protocol: Arbitrary Efficiency}\label{DBTProtocol}
\subsection{Setup}
We are now ready to describe a class of protocols that can remain secure at much lower BSM efficiencies. Additional to the previous assumptions, we also assume that all of the sources and Alice's local devices (as described below), do not have access to the routing choice of the switches. By shielding and spatially separating the multiple switches, we can also assume that each switch cannot learn the action of the other. Like before, we also assume unit detection efficiencies for the local measurement devices. We also consider the same BSM efficiency irrespective of the choice of local devices used for entanglement swapping. We introduce a technique called decoy Bell test (DBT), where with additional sources and devices, we also check the CHSH violation when $z=1$. In any given round, DBTs can occur with randomly selected devices.
\begin{figure}
    \centering
\includegraphics[width=0.9\linewidth]{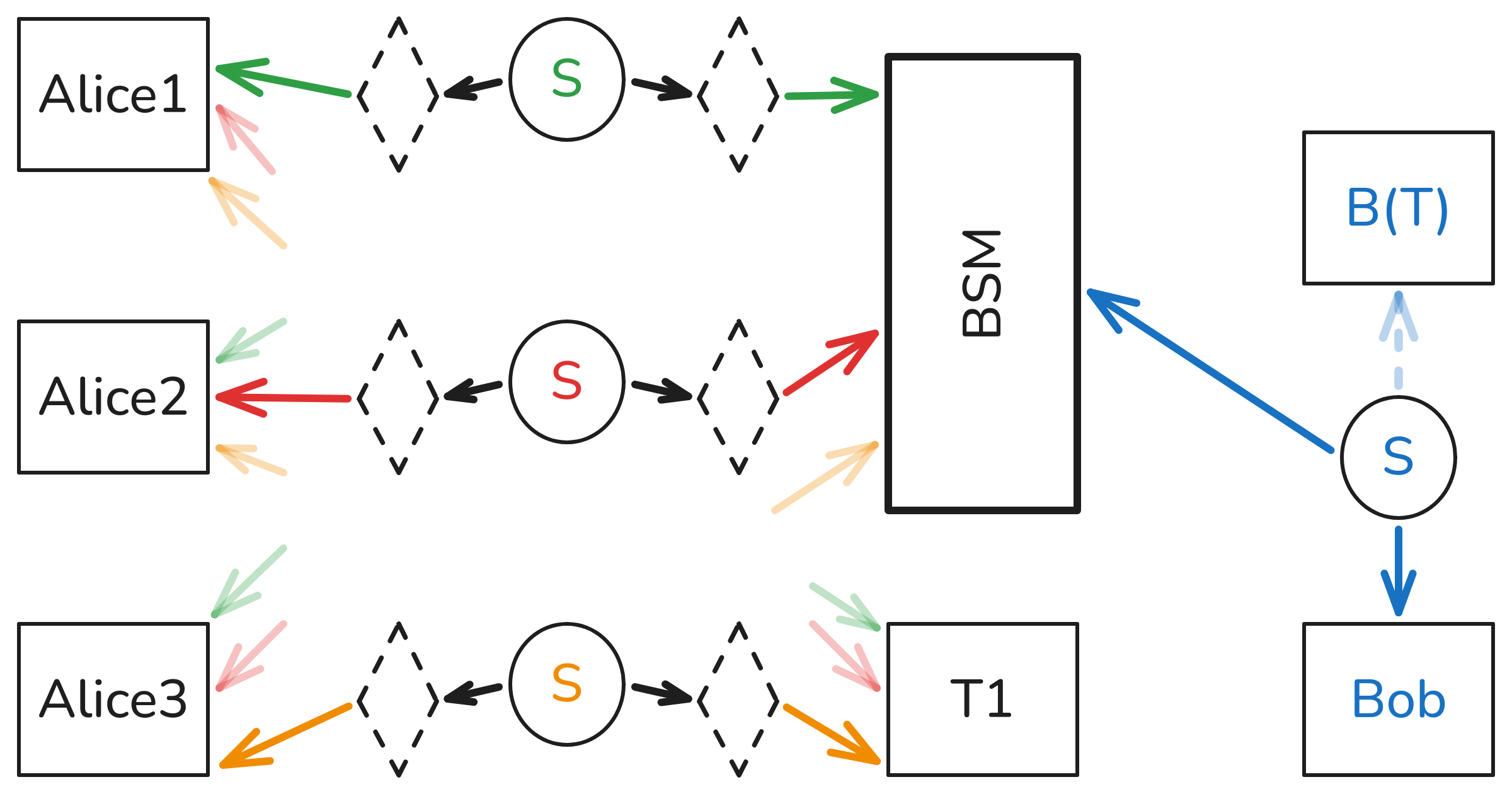}
    \caption{A decoy Bell test (DBT) is performed on Alice's side ($t=1$), in every round, including when $z=1$. In this figure, we show one instance of $z=1$, where the local test takes place with $A_3$ (Alice$3$), while other devices share their states with the $3-$port BSM ($N=3$). The semi-transparent and dashed arrows indicate the other possible routing paths, from the respective sources. Each switch is represented by a diamond with dashed lines. Two switches are placed alongside each source. The first switch routes to one of Alice's local devices randomly, whereas the second switch chooses one of the paths between the test device ($T_1$) and the BSM.}
    \label{fig:sbt}
\end{figure}
Consider an $N-$port BSM ($N>2$) with efficiency $\eta$, that takes in $N$ qubits from $(N-1)$ two-qubit sources on Alice's side, and from a single source like before on Bob's side. We can place Alice's local devices close to the BSM, and Bob farther away. Thereby, just one channel between Bob and the BSM may have intrinsically low transmission efficiency. Naturally, this construction can allow conference key agreement between multiple honest parties \cite{ribeiro2018fully}. However, we will stick to generating key between one of Alice's devices -- say Alice$1$ ($A_1$) -- and Bob ($B$). Alice's side has a total of $(N+t-1)$ sources, $(N+t-1)$ local devices, and $t$ test devices. We will denote the local devices (ideal inputs: $Z, X$) as: $A_1, A_2, ..., A_{N+t-1}$, and the test devices that are used solely for the local CHSH test as: $T_1, T_2, ..., T_t$. In each round, $t$ local tests are performed between $A_{l}$ and a test device, where $l \in \{1, 2, ...., N+t-1\}$. The remaining $(N-1)$ states are routed between the other devices and the BSM. All the switch inputs on Alice's side are controlled by the honest user $A_1$, which allows her to assign inputs to the nearby switches, such that each device receives one photon and the BSM receives $(N-1)$ photons.

The test device must not know the corresponding local device ($A_l$) involved in the CHSH test in the same round, to avoid loopholes. For this, we use two switches alongside each source. The first switch has $(N+t-1)$ possible routing choices towards one of Alice's local devices, and ensures that $A_l$ receives one photon in every round from a randomly selected source. The second switch has $(N+t-1)$ possible choices towards one of Alice's test devices, by which the other photon from the source is routed to any one of the $t$ test devices or to the $(N-1)$ BSM ports. As we are interested in generating key between $A_1$ and $B$, we can select all rounds where both of them use the same measurement inputs, and the other $(N+t-2)$ parties measure in the $X$ basis. 

We can now define the DBT condition to be satisfied for each of the devices on Alice's side. Let the overall average CHSH violation among all rounds for each device be $S_{A_{l}}=S$. For $z=1$ announcements (successful BSM rounds) at the $N-$port BSM, the conditional average CHSH violation that is observed,
\begin{equation}
    S_{A_{l}}^{z=1}=S_{A_{l}}=S \quad \forall l.
\end{equation}
If $S_{A_{l}}^{z=1} < S$, it would imply that the rounds with successful BSM projection have spuriously lower violation value than the overall average, and the rest of the rounds have a higher violation value. 

\subsection{Protocol}
We describe the protocol here, to establish the key between $A_1$ and $B$.

1. Each source is expected to ideally produce a Bell state in each round.

2. In every round, Bob ($B$) expects to receive one qubit from his nearest source. 

3. In every round, all local parties on Alice's side: $A_1,A_2, ..., A_{N+t-1}$, expect to receive a qubit from one of the $N+t-1$ sources on their side. A switch alongside each source randomises the qubit's destination of the local party.

4. The second qubit from each source is to be routed to the BSM or the test device(s) on the respective sides. On Alice's side, again, the qubit's destination is randomised using a second switch.

5. After all rounds, the BSM projection (failure or success, denoted by $z=0$ or $z=1$ respectively) results are announced. 

6. After this, the overall CHSH violation is estimated for each device. Also, the DBT condition is checked for $z=1$ events. If both are satisfactory, we proceed. Otherwise, we abort the protocol.

7. $A_1$ and $B$ announce their measurement inputs for all rounds.

8. Using classical communication, $A_1$ and $B$ estimate the QBER from a subset of $z=1$ rounds when they used the same bases ($x=y$). If they are within an acceptable threshold, they proceed with error correction and privacy amplification, for rounds where: $z=1$ when at least $A_1$ and $B$ share their states with the BSM (based on the routing path) and measure in the same bases, and at least one of the other local parties (on Alice's side) simultaneously shares its state with the BSM and measures in the $X-$basis.

\subsection{Security}
There are three sample events in the protocol based on the routing choices: the local test, failure of BSM projection ($z=0$) and successful BSM projection ($z=1$). The BSM announcement or the value of $z$, is correlated to the underlying source strategy, to minimize the violation corresponding to $A_1$ during key rounds (where $z=1$, and $A_1$ and $B$ are involved in the BSM projection). Since the local devices on Alice's side are agnostic to whether the state is routed to the test or the BSM, we can map out Eve's possible convex strategies for each device, to estimate the substantial violation. This is possible due to the routed local tests, where we can express $S$ as a linear combination of different weighted violations. Later, this substantial violation will be used to bound the conditional entropy, $H(A_1|E)$. For key rounds, the DBT condition mandates that the parties observe $S_{A_j}^{z=1}=S$, where $j \in \{2, 3, .., N+t-1\}$. Hence, Eve must assign $z=1$ accordingly. We describe a general strategy to identify the substantial violation for $z=1$ rounds.

We will consider the specific case where $N=3$. On Alice's side, we have local devices: $A_1, A_2, ..., A_{t+2}$. We can denote the substantial violation (unobserved) in key rounds as $S_{A_1}^{k}$. This is the term we want to minimize, and is equivalent to the substantial violation of Bob's device in the same rounds: $S_{B}^{k}$. We will refer to them as: $S_{A_1}^{k}=S_{B}^{k}=S_k$. In this setup, $A_1$ and $B$ can carry violations: $S_k$ or $2\sqrt{2}$, where $S_k\leq S$. Let $P$ be the probability with which $S_k$ occurs. Due to the DBT condition, not every degraded round can be accepted like before. Let $v$ be the probability of accepting a given degraded round as a key round (by assigning $z=1$). Then, the efficiency, $\eta=Pv$, and the overall violation for $A_1$ and $B$ is,
\begin{equation}\label{eqnv}
S=2\sqrt{2} \left( 1-\frac{\eta}{v}\right)+S_k\frac{\eta}{v}.    
\end{equation}
Our goal is to minimize $S_k$, which we will approach by maximizing $v$. 

Firstly, for attacks denoted by $i=0, 1, ..., t$, let attack $i$ degrade $A_1$, $B$ and $i+1$ other devices in a given round. These degraded devices carry states corresponding to violation $S_k$. The remaining devices carry $2\sqrt{2}$. For this attack, the probabilities $a_i$ and $b_i$ correspond to $z=1$ as described below. Next, we consider the case where only $A_1$ is degraded and all the remaining devices carry $2\sqrt{2}$. Here, $c$ is the corresponding probability for $z=1$ as described below. Let, 
\begin{itemize}
    \item $a_i$ be the probability of rounds where $z=1$ is assigned whenever any one degraded device apart from $A_1$ and $B$ lands in the BSM, 
    \item $b_i$ be the probability of rounds where $z=1$ is assigned like previously, and also when $A_1$ is sent to the test (latter requires $B$ and $2$ other degraded devices to land in the BSM (needs $i \geq 1$)),
    \item $c$ be the probability where $z=1$ is assigned whenever $A_1$ goes to the test, and any of the remaining devices land in the BSM.
\end{itemize}
Let $l_i$ be the number of $z=1$ routings whenever $A_1$ is sent to the test. This requires $2$ degraded states from $i+1$ devices to go to the BSM. Let $d(a_i)$ and $d(b_i)$ be the number of degraded devices tested within the attacks corresponding to $a_i$ and $b_i$ respectively. In the former, there are $(i+1)$ possible degraded devices (apart from $A_1$ and $B$), and $i$ devices go to the test in each of them. The latter includes the previous case, and also when $A_1$ is sent to the test at $\binom{i+1}{2}$ possible routings, we have $2$ degraded states at the BSM, leaving $(i-1)$ devices for the test. Hence,
\begin{align}
    l_i&=\binom{i+1}{2},\\
    d(a_i)&=i(i+1),\\
    d(b_i)&=i(i+1)+(i-1)\binom{i+1}{2}.
\end{align}
Then, we can write the following primal problem:
\begin{align}
  \max\quad & v=\sum_i\frac{i+1}{t'}\,(a_i+b_i),\label{v_exact}\\
  \text{s.t.}\quad & \sum_i(a_i+b_i)+c=1, \label{const1}\\
  & \sum_i l_i b_i+\binom{t'}{2}c=\binom{t'}{2}\eta,\label{const2}\\
  & \sum_i\bigl[d(a_i)a_i+d(b_i)b_i\bigr]\le t'\binom{t'}{2}\eta, \label{const3}\\
  & a_i,\,b_i,\,c\ \ge 0,
\end{align}
where $t'=t+1$ (number of devices on Alice's side apart from $A_1$) and all variables are non-negative. In the first equation, which we intend to maximize, we use the fact that $(i+1)/t'$ is the probability of $z=1$ among the attacks when $i+1$ devices are degraded. It is subject to three constraints. The first constraint \eqref{const1}, represents the overall fraction of attacks of $A_1$. The second constraint \eqref{const2}, represents the attacks within the DBT of $A_1$, where $\binom{t'}{2}$ is the number of routings where $A_1$ is tested. It is accepted at the rate of $\eta$. The third constraint, \eqref{const3}, bounds the attacks allowed by the DBTs of all the other devices on Alice's side. The right side of the inequality represents the maximum number of degraded test slots. The left side only includes the devices where $A_1$ is also degraded, hence the inequality. 

Let $\alpha, \beta$ and $\gamma$ be the prices or weights attached to the primal constraints \eqref{const1}, \eqref{const2} and \eqref{const3} respectively. For the dual problem, we write the minimization condition, followed by the conditions corresponding to the primal variables $a_i, b_i$ and $c$ respectively as
\begin{align}
   \min \quad & \alpha+\binom{t'}{2}\beta\eta+t'\binom{t'}{2}\gamma\eta, \\
   \text{s.t.} \quad & \alpha+\gamma i(i+1) \geq \frac{i+1}{t'}, \label{const2_1}\\ 
   &\alpha+\beta\binom{i+1}{2}  \notag \\
    &\quad +\gamma\left[ i(i+1) +(i-1)\binom{i+1}{2} \right] \geq \frac{i+1}{t'}, \label{const2_2}\\
    &\alpha+\beta\binom{t'}{2} \geq 0, \label{const2_3}
\end{align}
where $\gamma\geq 0$. With $a_i$ for example, $\alpha$ and $\gamma d(a_i)=\gamma i(i+1)$ are the price-weighted coefficients from \eqref{const1} and \eqref{const3}. The right sides of the inequalities are the coefficients of $a_i, b_i$ or $c$ in \eqref{v_exact}. By adding the constraints \eqref{const1}, \eqref{const2} and \eqref{const3}, with the prices as weights, we have the bound,
\begin{equation}\label{vbound}
    v \leq \alpha+\binom{t'}{2}\beta\eta+t'\binom{t'}{2}\gamma\eta,
\end{equation}
since the weighted sum of the left sides of the constraints is at least $v$. The dual problem minimizes the right side of the inequality in \eqref{vbound}. By strong duality, this minimum is equal to the maximum of $v$.

Along with $N=3$, we choose $t=2$, and obtain the maximum value of $v$ in each range of efficiencies (details in Appendix \ref{constraints}). This gives the minimum $S_{k}$ as a function of $\eta$, which is used to bound $H(A_1|E)$. The global QBER constrains the asymmetry in violations (proof in Appendix \ref{app: a}):
\begin{equation}
    h(q) \geq H(A_b| E) - H(A_1| E),
\end{equation}
where $A_b$ is the output with the maximum violation at the BSM port among $z=1$ rounds, that gives the minimum possible violation for the output $A_1$; and the observed $q$ is equivalent to $P(A_1\neq A_b)$. However, we will treat the minimum $S_k$ as in the zero QBER case. For finite QBER, $S_k$ would be lower, and the key rate would reduce. We ignore this effect when calculating $S_k$ in the rest of the work.
\begin{figure}
    \centering
\includegraphics[width= \linewidth]{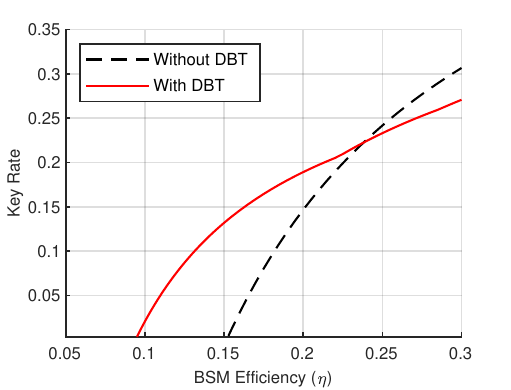}
    \caption{The key rate is plotted against the BSM efficiency at the minimum QBER, using honest Werner states with visibilities corresponding to $S=2.77$ for each local test, at NPA-level $4$ \cite{navascues2007bounding}. For the DBT case, a $3-$port BSM ($N=3$), and $2$ DBTs ($t=2$) on Alice's side are considered. DBTs enable key rates at much lower BSM efficiencies -- starting from $\eta \approx 9.4\%$ -- compared to $\eta \approx 15.2\%$ without DBTs.}
    \label{fig:R}
\end{figure}

In Figure \ref{fig:R}, we plot the key rate against $\eta$, by considering honest Werner states: $V~\ket{\psi^+}\bra{\psi^+}+(1-V)\mathbb{I}/4$, with {visibility} $V$ corresponding to $S=2.77$ for each local test, along with $N=3$ and $t=2$, at minimum $q$. The local Werner states severely penalize  higher $N$ alongside lower $S$ values, due to its high intrinsic $q$ with the global state's visibility: $V^N$. Without DBTs, increasing $N$ is not useful, and can always be associated with a two-port BSM. For the routed protocol without DBTs, the critical efficiency is around $15.2\%$. With DBTs, this comes down to $9.4\%$. Increasing $t$ does not necessarily lead to a bigger advantage. However, by increasing $N$, one can obtain positive key rates at arbitrary BSM efficiencies (at zero QBER). To see this, consider the $N-$port case where $t=1$. Alice's side has $N$ devices. We have three probabilities among the attacks: $a_0, a_1, b_1$. With $a_0$, $N-2$ devices on Alice's side (apart from $A_1$) are degraded. Since one device must go to the DBT, the acceptance rate is $1/(N-1)$. With $a_1$ and $b_1$, all devices are degraded, and all rounds are accepted. Therefore,
\begin{equation}
    v=\frac{a_0}{N-1}+a_1+b_1,
\end{equation}
which can be compared to the analogous equation in \eqref{v_exact} for $N=3$. Using \eqref{const1} and \eqref{const3}, we also have $a_0+a_1+b_1 \leq 1$ and $a_1+b_1\leq \eta$. Hence,
\begin{align}
    v&\leq\frac{1-a_1-b_1}{N-1}+a_1+b_1 \\
    &=\frac{1+(N-2)(a_1+b_1)}{N-1} \\
    &\leq \frac{1+\eta(N-2)}{N-1},\\
    &=\eta+\frac{1-\eta}{N-1},
\end{align}
and as $N\rightarrow\infty$, $v \rightarrow \eta$. Substituting this in \eqref{eqnv}, we get $S_k \rightarrow S$. Therefore, the substantial violation can theoretically approach $S$ for arbitrary $\eta$, as $N\rightarrow\infty$. This implies that a sufficient violation of the local tests can be a sufficient condition for arbitrary efficiencies. Still, very large values of $N$ are less feasible, and we may need to resort to better local tests with DBTs. For instance, critical efficiency of $1\%$ is possible at $S \approx 2.817$, with $N=3$ and $t=2$; compared to the critical efficiency of around $1.85\%$ without DBTs.

\begin{figure}[]
    \centering    \includegraphics[width=\linewidth]{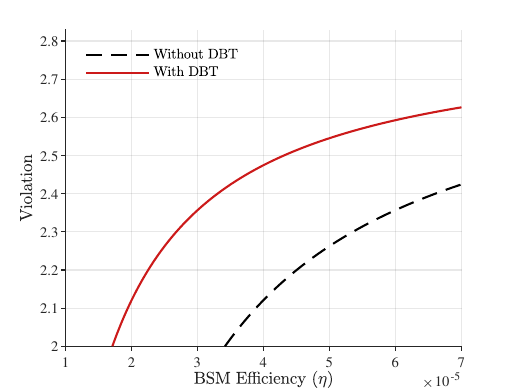}
    \caption{Substantial CHSH violation for $A_1$ ($S_k$) is plotted against $\eta$, for Werner states with very high visibility, $V=0.99999$. DBT ($N=3, t=1$) ensures better security at such ultra-low BSM efficiencies ($\sim 10^{-5}$).}
    \label{fig:sub_S}
\end{figure}

Pertaining to the repeated usage of $N=3$, we remark that an MDI-QKD protocol based on coincidence events at a $3-$port BSM, has been realized experimentally at a distance of 60 km \cite{yang2024experimental}. As $S$ increases to very high values, DBTs become less relevant, except for ultra-low efficiencies, like in existing BSM experiments (less than $10^{-5}$). The scaling with DBTs allows $S$ to be more robust in this case, as shown in Figure \ref{fig:sub_S}. A single DBT ($t=1$) is sufficient to observe an advantage at these efficiencies.
Following previous methods \cite{sekatski2025certification}, one can incorporate 
multiple inputs for each device to obtain stronger constraints. This can improve the local violation requirements, using the same techniques described in this section. However, using multiple inputs comes with its own set of challenges, separate from the standard CHSH experiment. The DBT technique may be integrated with such protocols with a multi-port BSM.

So far, we have assumed unit detection efficiency for the local devices. Let us briefly consider the case of finite detection efficiency and unit visibility (outside of no-click events). At limited detection efficiencies, to avoid loopholes, we can deterministically bin the outputs of rounds with no-click events to $0$, for each device. If we write, $S_k=2\sqrt{2}(\eta^L)^2+2(1-\eta^L)^2$, where $\eta^L$ is the detection efficiency of each local device, we require $\eta^L\gtrsim 92\%$ for the protocol to be secure. This is true for both cases: with and without DBTs. Highly efficient local devices are crucial for protocols with solely local Bell tests. The main difference is only in the fact that we have moved away from the long-range, to short-range tests.

\section{Routed Witnesses} \label{RoutedWitnesses}
\begin{figure}
    \centering
    \includegraphics[width=0.75\linewidth]{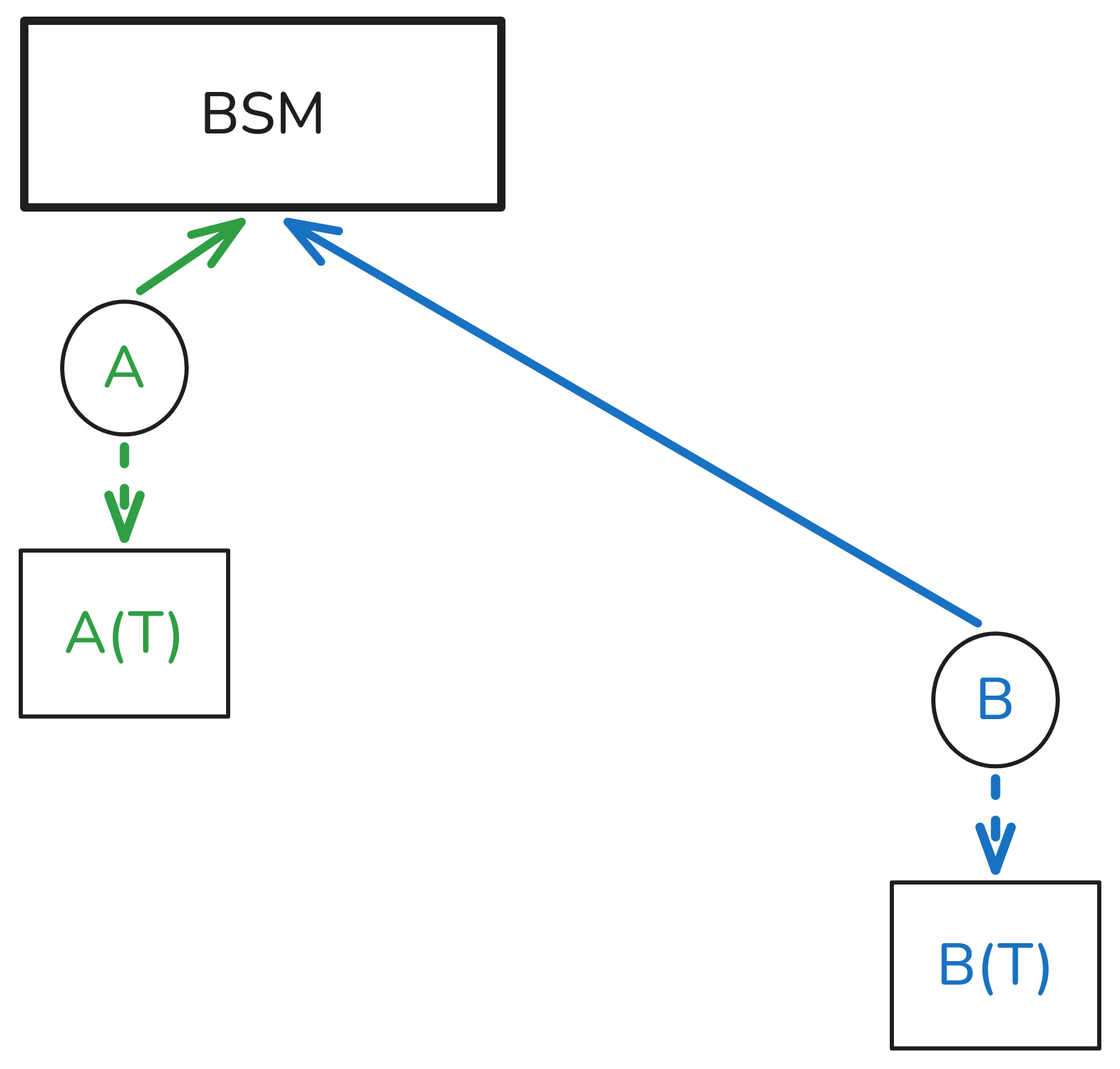}
    \caption{Alice and Bob have qubit-bounded sources, and use routed witnesses, similar to the routed Bell test scenario. The sources ($A$ and $B$) prepare and send qubits to the local measurement devices ($A(T)$ and $B(T)$) or to the distant BSM.}
    \label{fig:RW}
\end{figure}

We extend the concept of routing to the semi-device-independent (SDI) case with a key assumption about the source: the source is qubit-bounded and strictly emits one qubit in each round. The source is placed inside the lab. This protocol can also be seen as a modified version of measurement-device-independent quantum key distribution (MDI-QKD). We address this routed protocol because they are closely related to our previous scenario. Both protocols essentially transmit BB84 states to the BSM. As before, the switches on either side route the state to either destination for Alice and Bob as shown in Fig.~\ref{fig:RW}. The state is mostly sent to the distant BSM, but sometimes sent and measured close to the source. We also consider memoryless devices which do not share entanglement with any other device. 

Each party prepares and sends one of the four BB84 states in a round. The local tests use a two-dimensional witness \cite{gallego2010device} to certify the accuracy of the encoded states, $\ket{\psi_{xa}}$ for Alice and $\ket{\psi_{yb}}$ for Bob. It can be expressed as a CHSH-type test in the following form \cite{pawlowski2011semi, woodhead2015secrecy}
\begin{equation}
\label{routedtest}
S = \sum_{x,x',a,a'}(-1)^{xx'}[E(a=a'|xx')- E(a\neq a'|xx')] \leq 2,
\end{equation}
where $x'$ and $a'$ are the inputs and outputs of the local test device (respectively) on Alice's side, and the choice of $x$, $a$ and $x'$ are trusted and random. \Cref{routedtest} can be violated by quantum systems with Hilbert space dimension of $2$. The maximal violation of the witness is $S = 2\sqrt{2}$, with the measurement inputs of the local testing device being $(Z \pm X)/\sqrt{2}$, for perfectly encoded states $\ket \psi_{xa}$. If we obtain maximal violation with these inputs, the source must be emitting perfectly encoded BB84 states. We will not discuss the impact of decoy states, but it is vital to avoid photon number splitting attacks for the states sent to the BSM in a practical experiment.

In this SDI-QKD protocol, the same analysis for obtaining the substantial violation as the DI-QKD case holds. But for now, we will also assume that the two sources are independent of each other. This is reasonable because they are placed in well shielded labs, with which Eve cannot interact in real time. If $P_s$ is the probability with which each source carries a degraded violation: $S_k$, then $S=2\sqrt{2}(1-P_s)+S_kP_s$, where $s \in \{1, 2\}$. The independence condition causes the degraded violations to occur in independent rounds. At zero QBER, the key can be generated only when both devices carry the same violations, giving $P_1=P_2=P$ for the same observed $S$. Thus, $\eta=P^2$. The optimal attack reduces to an independent port model. Further, it can easily be extended to an $N-$port BSM, where all $N$ sources are independent. Here, $P=\sqrt[N]{\eta}$, and $S=2\sqrt{2}(1-\sqrt[N]{\eta})+S_k\sqrt[N]{\eta}$. The higher scaling significantly closes the gap between $S$ and $S_k$ for fixed $\eta$.

\section{Conclusion}

Routed Bell tests with BSMs have the advantage of enabling DI-QKD with solely local Bell tests. However, the observed violation may still be insufficient at low BSM efficiencies typical of long distances. As such, the advantage of the routed setup disappears, despite strong local tests. One alternative, common in current experiments, is to use event-ready setups with quantum memories, by delaying measurements until a successful BSM projection \cite{zapatero2023advances}. However, this has remained difficult, and may not be feasible for long distances. Quantum memories are an additional bottleneck.

This is addressed by the proposed modification of the DI-QKD protocol by including local decoy Bell tests (DBTs), where the devices are required to show a consistent violation, including when there is a successful projection at the BSM. This provides a significant advantage, reducing the critical efficiency requirements by up to $6\%$ in absolute terms, for $S \gtrsim 2.75$, with a $3-$port BSM. We show an example, at $S=2.77$, where the critical efficiency improves from around $15.2\%$, to $9.4\%$ using DBTs. One limitation in these values is that we do not consider the finite QBER to calculate the substantial violation. Incorporating this would reduce the key rate, particularly for lower visibilities, for the DBT case. For close to ideal visibilities, the advantage is markedly relevant at ultra-low efficiencies. These efficiencies are typical of current long-distance BSM experiments, and suggest that a single DBT can be useful despite the Bell test only containing very small imperfections. Further, we achieve arbitrarily small efficiencies using arbitrarily many ports.

DBTs may also be useful in other scenarios involving many distant parties, or in other complex networks. For instance, the sources may be replaced with those that produce GHZ states. They can also be integrated with other modified protocols which aim to improve the distance for DI-QKD protocols. In future work, we will look at other routed protocols within the BSM architecture. Experimentally, a feasible routing setup remains to be seen, and so are experiments with strong CHSH test violations, and high local detection efficiencies. However, by pushing the focus to short-range experiments, and removing the need for long-range Bell tests, several difficulties in DI-QKD experiments can be closed.

\acknowledgments
We thank E. P. Lobo for useful discussions and comments. SV is grateful to nodeQ for financial support. MP acknowledges funding from the “National Centre for
HPC, Big Data and Quantum Computing (HPC)”
Project CN00000013 HyQELM – SPOKE 10. MP
is grateful to the Royal Society Wolfson Fellowship
(RSWF/R3/183013), the Department for the Economy
of Northern Ireland under the US-Ireland R\&D Partnership Programme, the PNRR PE Italian National Quan-
tum Science and Technology Institute (PE0000023), and
the EU Horizon Europe EIC Pathfinder project QuCoM
(GA no. 10032223).
\bibliography{references}

\appendix

\section{Example}\label{constraints}

For $N=3$ and $t=2$, we have the following devices: $A_1, A_2, A_3, A_4$ and $B$. Here, $t'=3$, and \eqref{vbound} becomes: 
\begin{equation}\label{v_3}
   v \leq \alpha+3\beta\eta+9\gamma\eta. 
\end{equation}
Along with 
\begin{equation}\label{v_31}
    v=\frac{1}{3}a_0+\frac{2}{3}(a_1+b_1)+\frac{3}{3}(a_2+b_2),
\end{equation} 
from \eqref{v_exact}, we can write the conditions for the prices in Table \ref{tab:conditions} for each attack probability using \eqref{const2_1}--\eqref{const2_3}.

\begin{table}[H]
\centering
\small
\renewcommand{\arraystretch}{1.5}
\begin{tabular}{@{}cc@{}}
\toprule
Probabilities & Condition\\\midrule
$a_0$ & $\alpha\ge\frac13$\\
$a_1$ & $\alpha+2\gamma\ge\frac23$\\
$b_1$ & $\alpha+\beta+2\gamma\ge\frac23$\\
$a_2$ & $\alpha+6\gamma\ge1$\\
$b_2$ & $\alpha+3\beta+9\gamma\ge1$\\
$c$ & $\alpha+3\beta\ge0$\\
\bottomrule
\end{tabular}
\caption{Conditions on the prices for $t=2$.}
\label{tab:conditions}
\end{table}

Since each attack probability corresponds to a condition on the price, we can write the conditions with equality, for the corresponding range of efficiencies.

\begin{table}[H]
\centering
\small
\renewcommand{\arraystretch}{1.5}
\begin{tabular*}{0.4\textwidth}{@{\extracolsep{\fill}}ccc@{}}
\toprule
$\eta$ range & Probabilities & $(\alpha,\beta,\gamma)$\\\midrule
$[0,\frac29]$ & $a_0,\ a_1,\ b_1$ & $(\frac13,\ 0,\ \frac16)$\\
$[\frac29,\frac27]$ & $a_1,\ b_1,\ a_2$ & $(\frac12,\ 0,\ \frac1{12})$\\
$[\frac27,1]$ & $b_1,\ a_2,\ b_2$ & $(\frac35,\ -\frac1{15},\ \frac1{15})$\\
\bottomrule
\end{tabular*}
\caption{Optimal prices are stated for $t=2$. In each range, the conditions of Table~\ref{tab:conditions} for the listed probabilities are imposed with equality, giving the optimal $(\alpha,\beta,\gamma)$ that attains the bound in \eqref{v_3}.}
\label{tab:prices}
\end{table}

The optimal $(\alpha,\beta,\gamma)$ can be substituted in \eqref{v_3} to obtain the upper bound on $v$. Solving for the probabilities listed for each range in Table \ref{tab:prices} (other probabilities are zero) using \eqref{const1}--\eqref{const3} (\eqref{const3} as equality), and substituting them in \eqref{v_31}, gives the same value as obtained previously in \eqref{v_3}. This proves its tightness. By substituting this maximum value of $v$ in \eqref{eqnv}, we obtain the minimum value of $S_k$.

\section{QBER In Asymmetric Strategies}\label{app: a}

We know that
\begin{align}
    H(A_1, A_b|E) &= H(A_1 \oplus A_b, A_1|E)\\ 
    &\leq H(A_1 \oplus A_b|E) + H(A_1| E),
\end{align}
where we used subadditivity for the inequality. We also know that $H(A_b|E) \leq H(A_1, A_b|E)$. Hence
\begin{align}
    &H(A_b|E) \leq H(A_1 \oplus A_b|E) + H(A_1| E),\\
    &H(A_1 \oplus A_b|E) \geq H(A_b| E) - H(A_1| E).
\end{align}
We denote Eve's classical measurement as $M$. Then using the data processing inequality, $q=P(A_1\neq A_b)$, and the Fano's inequality, we have
\begin{align}
    &H(A_1 \oplus A_b|E) \leq H(A_1 \oplus A_b|M) \leq h(q),\\
    &\implies h(q) \geq H(A_b| E) - H(A_1| E).
\end{align}
\section{Hidden Variable Strategy}\label{app: b}

We will show that the previous proposals of uncharacterised qubits (without hidden variables) in MDI-QKD (no self-testing) \cite{yin2014mismatched} have zero key rate in the otherwise adversarial scenario with a hidden variable included on each side, $\lambda_A$ and $\lambda_B$. We only assume instead that the sources are independent. Consider that Alice's source (similarly for Bob's state $\rho_{B}$) emits the following pure states:
\begin{align}
    \rho_{A}^{\lambda_a}&=\ket{a}\bra{a} \\ \rho_{A}^{\lambda_{X_A}}&=\ket{X_A}\bra{X_A},
\end{align}
which can be seen as deterministically encoding the output index ($\lambda_A=a$) or XOR modulo two ($\lambda_A=X_A$) of input and output indices, $X_A = x \oplus a$, and similarly for Bob we have $Y_B = y \oplus b$. For zero QBER, we need to ensure that the following statistical constraints at the BSM are met:
\begin{align}
    p(z=1|x=y, a = b) &= 0, \\
    p(z=1|x=y, a \neq b) &= 0.5, \\
    p(z=1|x \neq y) &= 0.25,
\end{align}
where $z = 0, 1$ denotes the failure or success of the BSM projection respectively, and the maximum or ideal efficiency of the BSM is $50\%$. If the source only produces $\ket{a}\bra{a}$, they violate the mismatched bases constraints at low QBER. To see this, consider the example where Alice and Bob's states are $\ket{a}\bra{a}$ and $\ket{b}\bra{b}$ respectively, which Eve can easily learn by making a measurement in the known basis. Eve can set $z=0$ for $a=b$, and $z=0$ or $z=1$ randomly for $a \neq b$. The constraints for $x=y$ are satisfied. However, $ p(z=1|x \neq y, a = b) = 0$ and $p(z=1|x \neq y, a \neq b) = 0.5$, which violates the mismatched bases constraints. 

Additionally, however, if Alice and Bob's states are equally likely to be $\ket{X_A}\bra{X_A}$ and $\ket{Y_B}\bra{Y_B}$ respectively, Eve can set $z=0$ for $X_A = Y_B$, and $z=0$ or $z=1$ randomly for $X_A \neq Y_B$. This gives $ p(z=1|x \neq y, X_A = Y_B) = 0$ and $p(z=1|x \neq y, X_A \neq Y_B) = 0.5$. Note that $p(z=1|x \neq y, a = b) = p(z=1|x \neq y, X_A \neq Y_B)$ and $p(z=1|x \neq y, a \neq b) = p(z=1|x \neq y, X_A = Y_B)$. Therefore, by including the deterministic XOR states, we have $p(z=1|x \neq y) = 0.25$, which satisfies the mismatch bases constraint. Also, to satisfy the constraints, Eve will announce a failure when each state is encoded differently in the same round. In the absence of routed local tests, no key rate can be obtained. Like in the prepare-and-measure scenario, self-testing is mandatory for arbitrary qubit sources that may contain a hidden variable.
\section{Upper Bound}
\begin{figure}[htbp]
    \centering
\includegraphics[width=0.9\linewidth]{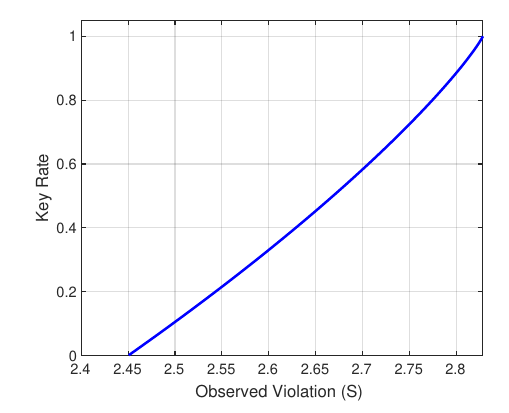}
    \caption{The upper bound on the key rate is plotted against observed $S$ at ideal efficiency ($\eta=1$), by considering Werner states for the local tests, at minimum QBER. It holds for both DI-QKD (without DBT) and the SDI protocol (without independence assumption between sources). For the DI-QKD protocol, the critical $S$ value (around $2.45$) is underestimated by approximately $1.6\%$, compared to using the lower bound.}
    \label{fig:Rw}
\end{figure}

We will use the upper bound based on the convex combination attack \cite{lukanowski2023upper}, using our hidden variable strategy, which may be of independent interest to other qubit certification scenarios. We will consider non-independent sources now, where $\lambda_A=\lambda_B$. This allows us to use the same analysis for both of the routed protocols that we discussed (without DBT). With the devices used to test the CHSH-type witness, we follow the same adversarial strategy of deterministically sending the output or XOR indices through each qubit to the self-test, for which $S=2$. This extremal local strategy corresponds to eight output combinations, $a_{x}b_{y}$, denoted by $\{a_{0}b_{0}$, $a_{0}b_{1}$, $a_{1}b_{0}$, $a_{1}b_{1}\}$, each activated during a specific round: $\{00, 00, 00, 00\}$, $\{00, 01, 00, 01\}$, $\{00, 00, 10, 10\}$, $\{01, 00, 11, 10\}$, and the remaining four are obtained by complementing all bits. For the four aforementioned combinations, notice that $a=0$ in the first and second, and $X_A=x \oplus a = 0$ in the other two. With one bit from the source, setting $b = a$ ($\lambda=a$) and $b = y$ ($\lambda=X_A$) respectively are sufficient for constructing extremal local correlations. The strategy is still constrained by observed $S$. Now, using a single source, we can write $S = 2\sqrt{2}(1-\eta)+S_{k} \eta$, where $S_k=2\sqrt{2}p^{NL}+2p^L$ is used for $z=1$, where $p^{NL}$ and $p^{L}$ are the proportion of ideal-violation strategy ($S_k=2\sqrt{2}$) and the extremal local strategy ($S_k=2$) respectively, and $p^{NL}+p^{L}=1$. For this individual attack, no information can be obtained when $S_k=2\sqrt{2}$. The upper bound on the key rate, 
\begin{equation}
    R_u\leq H(A|E)-H(A|B)=p^{NL}-h(q).
\end{equation}
This is also a simpler alternative applicable for DI-QKD, and known to give threshold values close to the analytic lower bound \cite{lukanowski2023upper}. No security is possible when $\eta\leq p^L$. When $\eta > p^L$, we can write $S_k=2\sqrt{2}p^{NL'}$, and $q=0.5(1-(p^{NL'})^2)$, considering depolarizing noise. We plot the key rate for ideal efficiency in Figure \ref{fig:Rw}.
\end{document}